\documentclass[journal,9pt]{IEEEtran}
\usepackage{amsmath,amsfonts}
\usepackage{algorithmic}
\usepackage{algorithm}
\usepackage{array}

\usepackage[caption=false,font=normalsize,labelfont=sf,textfont=sf]{subfig}
\usepackage{textcomp}
\usepackage{stfloats}
\usepackage{url}
\usepackage{verbatim}
\usepackage{graphicx}
\usepackage{cite}
\usepackage{hyperref}
\usepackage{placeins}
\usepackage{color,soul}
\begin{document}
\bstctlcite{IEEEexample:BSTcontrol}

\title{\huge{Equivalent Circuit Modeling of a Lumped-element Loaded Metasurface under Arbitrary Incidence and Polarization}}

\author{\Large{
Athanasios Nousiou, 
Christos K. Liaskos,
Nikolaos V. Kantartzis, and Alexandros Pitilakis,
}

%\author{\Large{Athanasios Nousiou, Nikolaos V. Kantartzis, and Alexandros Pitilakis}
        % <-this % stops a space
%\thanks{This paper was produced by the IEEE Publication Technology Group. They are in Piscataway, NJ.}% <-this % stops a space
\thanks{Manuscript submitted April 2025.
(\textit{Corresponding author: Alexandros Pitilakis.}) }%
\thanks{A. Nousiou, N.V. Kantartzis and A. Pitilakis are with the School of Electrical and Computer Engineering, Aristotle University of Thessaloniki, 54124 Greece (email: athanous@ece.auth.gr; kant@auth.gr; alexpiti@auth.gr).}
\thanks{C.K. Liaskos is with the Computer Science Engineering Department, University of Ioannina, Ioannina, and Foundation for Research and Technology Hellas (FORTH), Greece (cliaskos@ics.forth.gr).}
}

%\markboth{IEEE Transactions on Antennas and Propagation,~Vol.~8, No. 23, 2023}%
%\markboth{IEEE Transactions on Antennas and Propagation}{Nousiou, Pitilakis and Kantartzis: TITLE}
\markboth{Submitted for publication}{Submitted for publication}

%\IEEEpubid{0000--0000~\copyright~2025 IEEE}

\maketitle

\begin{abstract}
%In recent years, Reconfigurable Intelligent Surfaces (RIS) have emerged as a key technology for next-generation communication, offering real-time and dynamic control over electromagnetic wave characteristics. The scattering of an RIS unit cell can be effectively modeled using an equivalent circuit model (ECM) of a lumped RLC-loaded metasurface. However, existing ECMs, particularly those with square patches, often fail to generalize beyond specific configurations or integrate lumped elements, limiting their practical applicability to single polarization or normal incidence.
In this work we develop a simple equivalent circuit model (ECM) that predicts the spectral response of a lumped-element loaded single patterned layer reflective metasurface. The proposed ECM maintains accuracy across varying design parameters including dual-polarization, arbitrary obliquity and azimuthal rotation of incidence plane, as well as physical dimensions of the lumped elements. Our approach is built incrementally, starting from an elementary unit cell which is extended to a 2-by-2 arrangement that allows for multifunctionality and reconfigurability in both polarizations, independently. These capabilities facilitate the design of tunable holographic metasurfaces which can implement reconfigurable intelligent surfaces (RIS) for advanced wireless communication systems. We evaluate the limitations and accuracy of the proposed ECM with full-wave simulations (FWS), which evidence robustness up to 45 degrees and thus highlight its aptitude as a rapid design tool or a guide for FWS-based optimization.
\end{abstract}


% ==========================================================
\section{Introduction}
\label{sec:1:introduction}
% ==========================================================
\IEEEPARstart{M}{etasurfaces} are engineered 2D materials composed of subwavelength-scale particles that enable precise control over scattered electromagnetic wave properties such as magnitude, phase, and polarization \cite{Review1,Review2}. In recent years, the RIS paradigm has emerged, heralding that dynamic manipulation of scattering from reconfigurable metasurfaces can enable smart antennas or programmable environments that  open new dimensions for enhanced wireless communications \cite{Cui2014,danufane2021on,liaskos2022xrrf}.The RIS versatility is realized by electrically controlled elements, such as diodes or varactors. This technology is central to next-generation microwave, millimeter-wave and THz wireless communications systems, like 6G, addressing traditional system limitations by improving performance, energy efficiency, and cost-effectiveness \cite{Zhang2021,Zhu2023}. Beyond telecommunications, metasurfaces have applications in sensing, imaging, holography, and stealth.  

Despite the disruptive potential of RIS, the design and analysis of metasurface unit cells is still hampered by resource-demanding FWS \cite{Pitilakis2021} which is a considerable bottleneck in the fast-paced development and deployment of wireless communication hardware and infrastructure. Thus, an outstanding challenge is the development of analytical models that can accurately predict the metasurface response or that can guide initial parameter optimization, thus minimizing the required FWS fine-tuning; these models should operate consistently across a wide parameter space and should also provide insight into physical limitations. Existing models \cite{Sievenpiper2003,Costa2010,Lopez2022,PrezEscribano2024,Ma2024}, apart from being structure-specific, are often limited to static configurations, to normal incidence, to a single polarization, or to only one of the principal Cartesian planes; all these aspects restrict their applicability to the broad range of dynamic real-world scenarios.

This work addresses these shortcomings by introducing a simplified yet comprehensive analytical ECM for metasurfaces composed of square patch grids with lumped elements `loading' the gaps, accommodating both orthogonal linear polarizations at oblique directions or when the incidence plane is azimuthally rotated with respect to the grid. The ECM builds upon transmission line (TL) theory and incorporates the dispersive surface impedance of the patterned and loaded metallic layers, systematically evolving from fundamental ($1\times1$ cell \cite{Liu:2019}) to advanced ($2\times2$ cell \cite{pitilakis2023reconfigurable}) structures, without relying on parameter retrieval or fitting from measurements or FWS and circuit simulators. Our model integrates key parameters such as unit cell period, substrate properties, patch gaps, lumped element values (resistance and capacitance) and physical dimensions, incidence angle, and polarization. Balancing simplicity and accuracy, it maintains a resonance-frequency error below {$15\%$} between analytical and simulated results {which holds for oblique incidence angles up to $45^\circ$, corresponding to the practical range in most applications}. Given that RIS are dynamically tunable, i.e., such minor deviations in frequency or performance can be compensated post-fabrication, these versatile unit cell architectures can provide support multifunctionality (simultaneous amplitude and phase control), reconfigurability (independent tuning of unit cells), and dual-polarization (independent responses along the $x$ and $y$ axes). These features pave the way for the streamlined design of holographic RIS \cite{Pitilakis2024} for advanced wireless applications. Finally, this model lays the groundwork for developing more generalized ECMs, potentially independent of specific grid configurations or incorporating more complicated structures, such as multiple patterned layers, through-vias, and magnetic resonances.

The paper is structured as follows: Following this introduction, Section~\ref{sec:arch_meth} presents the unit cell architectures considered in this work and outlines the analytical tools used to model their response in progressively more complex configurations. Section~\ref{sec:Results} contains the evaluation of the ECM by means of comparison with FWS predicted responses for various metasurface functionalities. Section~\ref{sec:4:Future} holds the conclusion and future prospects of this work.

%\IEEEpubidadjcol % see line with "\IEEEpubid{0000--0000~\copyright~2025 IEEE}" % ---> Remember, if you use this you must call \IEEEpubidadjcol in the second column for its text to clear the IEEEpubid mark.

% ==========================================================
\section{Architecture and Methodology}
\label{sec:arch_meth}
% ==========================================================

% ----------------------------------
\subsection{Unit Cell Architecture} \label{sec:arch}

\begin{figure}[!t]
    \centerline{\includegraphics[width=85mm]{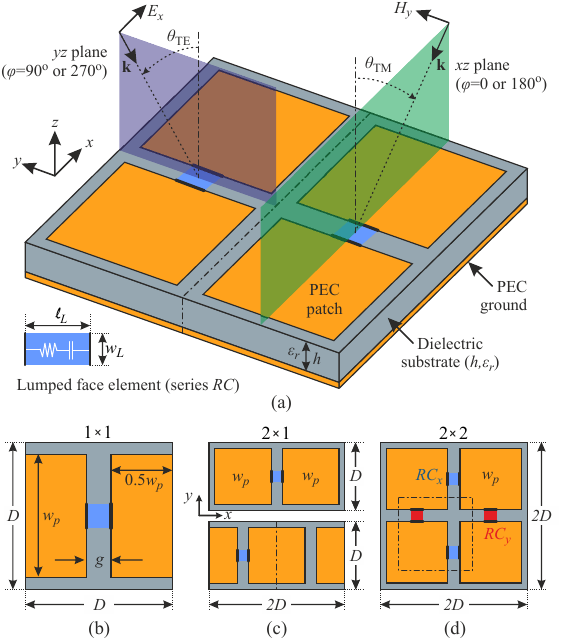}}
    \caption{(a) Definition of unit cell architecture, principal planes and polarizations. (b)-(d) Top-view evolution of the $1\times1$, $2\times1$, and $2\times2$ unit cell configurations. The lumped loads are modeled as `face elements' (ribbons) of given length and width and given series RC values.}
    \label{fig:Geoms}
\end{figure}

The reflective metasurfaces studied in this work consist of a periodic repetition of subwavelength elements on a grounded dielectric substrate, where geometrical shapes (e.g., squares, crosses, rings) determine electromagnetic wave manipulation. By selecting element shapes, arrangement, and substrate properties, metasurfaces control the scattered wave magnitude and phase. Similar control is achievable by integrating \textit{tunable} elements, like varactors or varistors \cite{Pitilakis2021}, while maintaining the geometry unchanged across the whole aperture. Some simple yet flexible designs, see Fig.~\ref{fig:Geoms}, include unit cells with the following configurations:
\begin{itemize}
    \item \textbf{$1\times1$}, Fig.~\ref{fig:Geoms}(b): Square patch grid connected by lumped elements in one dimension. Adding a lumped load also in the other direction, dash-dot outlined rectangle in Fig.~\ref{fig:Geoms}(d), allows for dual-polarization control but only for global tuning; for individual tuning, cross-talk between adjacent cells, i.e., non-local effects, must be accounted for.
    \item  \textbf{$2\times1$}, Fig.~\ref{fig:Geoms}(c): Pair of patches connected by a lumped element in one dimension; minimizes aforementioned cross-talk but affords control in one plane only.
    \item  \textbf{$2\times2$}, Fig.~\ref{fig:Geoms}(d): Extension of the $2\times1$ configuration in two dimensions, which allows for independent control in two planes; the parallel pair of loads, $RC_x$ and $RC_y$, are assumed identical in this work.
\end{itemize}

By tuning the resistance and capacitance of the lumped loads integrated in each unit cell, a large range of amplitude and phase tuning can be achieved, enabling real-time dynamic control. This localized tuning allows finite-aperture metasurfaces to shape reflected wavefronts for applications like beam steering, focusing, or holography. In RF, microwave, or millimeter-wave metasurface designs, such reconfigurability can be implemented electrically using tunable varistors and varactors. Additionally, incorporating different lumped elements along the two lateral axes, $x$ and $y$ in Fig.~\ref{fig:Geoms}, enables independent control of the two linear polarizations in $1\times1$ and $2\times2$ cell configurations. Simple functionalities like plane-wave absorption and polarization conversion can be accomplished by global tuning, whereas more complex ones, such as wavefront shaping, require localized tuning and clustering into supercells \cite{Liu:2019}. This work focuses on square elements for their isotropic response and the simplicity in their modeling and assumes all metals are perfect electrical conductors (PEC) with zero thickness, for further simplification.

% ----------------------------------
\subsection{Transmission Line Modeling} \label{sec:TLM}

The design process begins with theoretical modeling leveraging ECM and TL modeling (TLM) approaches \cite{Sievenpiper2003,Luukkonen2009,Costa2010,Lopez2022,PrezEscribano2024,Ma2024,Pitilakis2024}. The ECM/TLM can be used to rapidly \textit{analyze} a given cell, i.e., to quantitatively compute its co/cross-polarized reflection coefficient spectra or predict how each physical parameter qualitatively affects the response. Alternatively, it can be easily `inverted' to \textit{design} a unit cell, i.e., to select its physical parameters (period, patch gaps, substrate material and thickness, RC loadin, etc.) given a desired response to a given incident plane wave (frequency, direction, polarization).
%In this case, patch size is chosen to align with the operating frequency, substrate properties are adjusted for optimal resonance and, finally, resistance and capacitance ranges are selected for accurate frequency tuning. 
In both workflows, analysis or design, FWS using established commercial software (e.g., CST, HFSS, or COMSOL) \cite{Pitilakis2021} can be used to refine and validate the design. 

\begin{figure}[!t]
    \centerline{\includegraphics{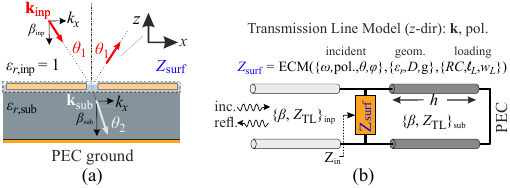}}
    \caption{(a) Oblique incidence on the unit cell: the refracted angle $\theta_{2}$ is computed by Snell's law and is equal for TE and TM polarizations, assuming isotropic bulk substrates. (b) Transmission line modeling for the cell reflection; the loaded patch grid is represented by an ECM-derived $Z_\text{surf}$ while the TL characteristic impedance and propagation constant depend on polarization and the $\textbf{k}$ in the medium.}
    \label{fig:TLM}
\end{figure}

The chosen unit cell architecture can be efficiently modeled using TLM: Dielectric slabs are modeled as short TL segments whose characteristic impedance and propagation constant depend on their permittivity and on the incident wavevector $\mathbf{k}$ and polarization, decomposed in transverse electric (TE) and transverse magnetic (TM) planes; note that $\mathbf{k}$ incorporates the information of the plane wave direction in 3D space and the harmonic frequency. Patterned conductor layers are modeled by an equivalent complex surface admittance $Y_\text{surf}$ or impedance $Z_\text{surf}=1/Y_\text{surf}$, which is computed by approximate dispersive ECMs\cite{Luukkonen2009}; ground planes are equivalent to short circuits. This modeling approach is depicted in Fig.~\ref{fig:TLM} and its accuracy is very good for substrates with electric thickness above $\lambda/4$; below that limit near-field coupling \cite{POMC} progressively diminishes the validity of the ECM used for $Z_\text{surf}$. 

The input impedance of the unit cell, $Z_\text{in}$ in Fig.~\ref{fig:TLM}(b), is a function of various parameters including the operating frequency, angle of incidence, polarization, geometry and lumped loading. Knowing $Z_\text{in}$, the reflection coefficient can be calculated 
 \begin{equation} \label{eq:GammaTL}
     \Gamma = \frac{Z_{\mathrm{in}}- Z_\text{TL,inp}}{Z_{\mathrm{in}} + Z_\text{TL,inp}}=
     \frac{Y_\text{TL,inp}-Y_{\mathrm{in}}}{Y_\text{TL,inp} + Y_{\mathrm{in}}},
 \end{equation} 
where $Z_\text{TL,inp}$ is the characteristic impedance of the TL that feeds the structure. For a plane wave traveling inside an $\varepsilon_r$ slab of a stratified structure, obliquely at angle $\theta$ with respect to the $z$-axis,
 \begin{equation} \label{eq:ZTLslab}
     Z_\text{TL} = \begin{cases}
         \frac{\eta_0}{\sqrt{\varepsilon_r}} \cos{\theta},\space &\text{TM pol.} \\
         \frac{\eta_0}{\sqrt{\varepsilon_r}}\sec{\theta},\space &\text{TE pol.}
     \end{cases}
 \end{equation}
where $\eta_0\approx377~\Omega$ is the characteristic impedance of vacuum; in normal incidence, $Z_\text{TL}$ matches the bulk medium wave impedance; for lossy dielectrics, the $\text{Imag}\{\varepsilon_r\}$ leads to absorption.

In the absence of the patterned metallic layer ($Z_\text{surf}=\infty$ in Fig.~\ref{fig:TLM}), the input impedance is simply the short-circuit (ground-plane) transformed by an $h$-long TL, where $h$ is the substrate thickness, i.e., 
\begin{equation} \label{eq:Zin_ShortedTL}
    Z_\text{in} = Z_\text{sub,GND}=jZ_\text{TL,sub} \tan(\beta_\text{TL,sub} h).
\end{equation}
In this expression, $\beta_\text{TL,sub}= \sqrt{k_\text{sub}^2 - k_\text{tan}^2}$ is the propagation constant along the TL, where $k_\text{sub}=k_0\sqrt{\varepsilon_{r,\text{sub}}}$ and $k_\text{tan}= |\mathbf{k}_\text{inp}| \sin \theta_1 = |\mathbf{k}_\text{sub}| \sin \theta_2$; the latter is Snell's law, which can be rewritten in its most usual form as $\sqrt{\varepsilon_{r,\text{inp}}}\sin\theta_1 = \sqrt{\varepsilon_{r,\text{sub}}}\sin\theta_2$; in this framework it is used to compute the plane wave direction $\theta_2$ in the substrate when the incidence angle $\theta_1$ and the refractive indices of the cladding and substrate are known.

In the presence of a single patterned metallic layer on the input port, i.e., on the interface of a semi-infinite `input' dielectric and the substrate, Fig.~\ref{fig:TLM}, the input admittance of the whole unit cell is given by the parallel combination
\begin{equation} 
    Y_\text{in} = Y_\text{surf} + Y_\text{sub,GND}.
\end{equation}
In this expression, $Y_\text{surf}$ can be highly dispersive on geometry, materials, and incident wave properties (frequency, direction, polarization).

%-----------------------------------------------------------
\subsection{ECM for Unloaded Patch-Grid} \label{sec3:ECM_unloaded}

The most crucial part in the above methodology is the broadband surface impedance of the patterned metallic layer, $Z_\text{surf}$. When simulated or measured data are available, retrieval methods can be used \cite{Nitas2022}. In this work, we aim to develop a simple yet accurate enough analytical ECM for the surface impedance, which can then guide FWS-based optimization, if needed. Unless otherwise stated, the plane of incidence required to distinguish the two orthogonal linear polarizations (TE and TM) matches one of the two principal planes, i.e., $xz$ or $yz$, as defined in Fig.~\ref{fig:Geoms}(a).

Starting from the $1\times1$ cell, Fig.~\ref{fig:Geoms}(b), we note that in the absence of loading (e.g., $R_\text{series}\rightarrow\infty$), the dispersive patch-grid impedance can be computed by the high-impedance surface ECM developed in \cite{Lukkonen2008}: When the patch grid is embedded within two not-too-different different and not-too-thin dielectric media, we can consider an equivalent bulk medium of effective permittivity given by $\varepsilon_{r,\text{eff}} = (\varepsilon_{r,1} + \varepsilon_{r,2})/2$, where $\varepsilon_{r,1/2}$ are bulk permittivities of the two media, e.g., the air cladding and the dielectric substrate in the cases of Fig.~\ref{fig:Geoms}. For narrow gaps, the surface impedance of the unloaded patch grid (UPG) grid is capacitive,
\begin{equation} \label{eq:Z_Luukk}
    Z_\text{UPG} = \begin{cases}
        -j\frac{\eta_{0}/\sqrt{\varepsilon_{r,\mathrm{eff}}}}{2\alpha}, &\text{TM pol.} \\
        -j\frac{\eta_{0}/\sqrt{\varepsilon_{r,\mathrm{eff}}}}{2\alpha \left(1-\frac{1}{\varepsilon_{r,\mathrm{eff}}}\frac{\sin^2{\theta_\text{inc}}}{2} \right)}, &\text{TE pol.}
    \end{cases}
\end{equation}
where $\alpha$ is the `grid parameter' approximated by
\begin{equation} 
    \alpha = \frac{k_\text{eff}D}{\pi} \ln{\bigg({\csc{\bigl(\frac{\pi g}{2D}\bigr)}}\bigg)};
\end{equation}
in this expression, $D$ is the unit cell pitch (period), $g$ is the gap between the patches, and $k_\text{eff} = k_0\sqrt{\varepsilon_{r,\text{eff}}}$. Note that the dispersive $Z_\text{surf}$ of Eq.~\eqref{eq:Z_Luukk} accurately captures only the primary (lowest frequency) resonance, which is nevertheless the most important; higher order modes can be accounted for with more complicated formulas, e.g., Section~5.18 in \cite{Marcuvitz1951}.

The capacitive $Z_\text{UPG}$ and the inductive $Y_\text{sub,GND}$ give rise to an RLC-like circuit adequately described by a Lorentzian lineshape, i.e., quantified by a resonance frequency and a finite quality factor if there are losses in the system, e.g., substrate $\tan\delta\neq0$. 
%This formula can be used to optimize the geometry $\{D,s,h\}$ for target performance in terms of $\Gamma$ at given $\{f,\theta_1,\text{pol.}\}$, or vice-versa.

%Both Eq. (\ref{X02}) and (\ref{X03}) are derived using the methodology described in \cite{Lukkonen2008} and the parallel combination of them will lead to equations in \cite{Luukkonen2009}. Additionally, we can note that both of these parameters depend on the polarization state of the incident wave, under oblique incidence. 

%$\beta$ is the propagation constant of the equivalent TL, $k_{1/2}$ are the bulk wave-numbers of the cladding/substrate and cladding media, 

%-----------------------------------------------------------
\subsection{ECM for the $1\times1$ Cell} \label{sec3:ECM_1x1}

To realistically model lumped loads such as surface-mount device (SMD) capacitors and resistors, in the unit cell architectures in Fig.~\ref{fig:Geoms} we consider lumped face elements (LFE), i.e., zero-thickness ribbons with length $\ell_L=g$ (equal to the patch gap) and nonzero width $w_L$. Following \cite{Tret}, we found that the surface admittance for such a loaded-grid cell can be approximated by the parallel connection:
\begin{equation} \label{eq:Zsurf_1x1}
    Y_\text{surf}^{(1\times1)} = Y_\text{UPG} + Y_\text{LFE}^{(1\times1)} 
\end{equation}
where the $Y_\text{UPG}=1/Z_\text{UPG}$ directly from Eq.~\eqref{eq:Z_Luukk} and $Y_\text{LFE}^{(1\times1)}$ is an equivalent admittance that captures both for the LFE's physical dimensions and its lumped impedance, e.g., the complex number $Z_\text{load}=R_\text{series}+1/(j\omega C_\text{series})$.

For very wide LFE, $w_L\rightarrow w_p=D-g$, it was found that $Y_\text{LFE}^{(1\times1)}\rightarrow 1/Z_\text{load}$, i.e., the LFE physical dimensions have negligible contribution. However, for narrower LFEs, near the 2:1 (length:width) form-factor of commercial off-the-shelf SMDs, a correction term is needed, e.g., in series
\begin{equation} \label{eq:ZLFE_1x1}  
    Z_\text{LFE}^{(1\times1)} \approx Z_\text{load} + Z_\text{corr}. 
\end{equation}  
For $Z_\text{corr}$, we found very good agreement with the reactive part of the impedance related to the width-discontinuity along a microstrip (MS) TL: Replacing the LFE by a narrow ribbon, we have a discontinuous MS TL along the $x$-axis in topview Fig.~\ref{fig:Geoms}(b), so we can compute two MS characteristic impedances, $Z_{0,w_p}$ for the patches and $Z_{0,w_L}$ for the ribbon, using standard textbook formulas \cite{pozar}. Thus, the correction term can be approximated by 
\begin{equation}  
 \label{eq:Zcorr1x1}  
 Z_\text{corr} \approx j\mathrm{Imag} 
 \biggl\{  Z_{0,w_L} 
 \frac{Z_{0,w_p} + jZ_{0,w_L} \tan{(\beta_\text{LFE}}\ell_L)}
      {Z_{0,w_L} + jZ_{0,w_p} \tan{(\beta_\text{LFE}}\ell_L)}
 \biggr\}  
\end{equation}  
where $\ell_L=g$ is the LFE length (equal to the patch-gap), and $\beta_\text{LFE}$ is the propagation constant for a MS with the LFE’s width. From Eq.~\eqref{eq:Zcorr1x1}, it is evident that $Z_\text{corr}$ vanishes for very narrow patch gaps ($\ell_L\rightarrow0$) as well as for $w_L\rightarrow w_p$, as mentioned. Finally, we note that Eq.~\eqref{eq:Zcorr1x1} is dispersive only on frequency and structure properties (dimensions and materials) while its dispersion on obliquity angle and polarization is omitted in this study, for the sake of simplicity. 

%-----------------------------------------------------------
\subsection{ECM for the $2\times1$ and $2\times2$ Cells} \label{sec3:ECM_2x1}

The $2\times1$ cell can be seen as a combination of two $1\times1$ cells, only one of which has its gap loaded by a lumped element, bottom panel in Fig.~\ref{fig:Geoms}(c). For its surface admittance we initially tested the simple average of the admittances of the two sub-cells \cite{Pitilakis2024}, i.e., Eqs.~\eqref{eq:Z_Luukk} and \eqref{eq:Zsurf_1x1}, which suffered from limited accuracy or validity range, in frequency and/or obliquity. Our proposed approach is using an equivalent $Y_\text{LFE}^{(2\times1)}$ that functions similarly to the $1\times1$ case, i.e., $Y_\text{surf}^{(2\times1)} = Y_\text{UPG} + Y_\text{LFE}^{(2\times1)}$. The corresponding equivalent impedance can be written as the parallel combination of two equal impedances, assuming the overall effect is evenly (homogeneously) distributed on the cell:
\begin{equation} \label{X3}
    Z_\text{LFE}^{(2\times1)} \approx 2 \Big( Z_\text{LFE}^{(1\times1)} + Z_\text{cpl} \Big).
\end{equation}
We found that this particular representation encapsulates both the impedance contribution of the lumped RC element with its physical dimensions as well as the `nonlocal' coupling between loaded and unloaded $1\times1$ cells, in terms $Z_\text{LFE}^{(1\times1)}$ and $Z_\text{cpl}$, respectively. Term $Z_\text{LFE}^{(1\times1)}$ is taken from Eq.~\eqref{eq:ZLFE_1x1} and $Z_\text{cpl}$ can, in the presence of the ground plane, be quantified by means of the grounded slotline waveguide under the transverse resonance technique \cite{Luukkonen2009}; technical details can be found in the Appendix. 
%It is important to note that $Z_\text{LFE}^{(2\times1)}$ is an \textit{equivalent} lumped element that models the behavior of the component and the coupling between two adjacent patches in this structure. When placed in a $1 \times 1$ structure, it produces the same response as in the $2 \times 1$ configuration. 
Lastly, note that the geometry exhibits two axes of anisotropy; however, the axis without lumped elements behaves as an unloaded grid.

Concerning the $2 \times 2$ cell, Fig.~\ref{fig:Geoms}(d), formed by a concatenation of two $2\times1$ cells: Its response in the two principal planes ($xz$ and $yz$) can be extracted from the `isolated' $2\times1$ cell responses, assuming the pair of parallel loads along each transverse Cartesian axis has identical RC values, i.e., $RC_x^\text{top}\equiv RC_x^\text{bottom}$ and $RC_y^\text{left}\equiv RC_y^\text{right}$, the  for blue- and red-colored LFEs in Fig.~\ref{fig:Geoms}(d), respectively. Evidently, by assigning different $RC_{x/y}$ values, we can control the response in each plane independently, allowing for dual polarization operation.

%-----------------------------------------------------------
\subsection{Response under Arbitrary Plane of Incidence} \label{sec:arbPhi}

Note that when the plane of incidence is $xz$ ($\varphi=m\pi$, where $m$ is an integer) or $yz$ [$\varphi=(2m+1)\pi/2$] we expect no cross-polarized scattering under linearly polarized illumination, for all $\theta$, due to the rectangular geometry of the metal patterning, Fig.~\ref{fig:Geoms}(a). In this section, we derive an analytic expansion of our ECM/TLM to compute the unit cell response under oblique incidence with arbitrary $\varphi$ and for both linear polarizations, $s$ (TE) and $p$ (TM). The unit cells studied in this work exhibit two orthogonal axes of anisotropy, aligned with the $x$ and $y$ axes, 
%This approach is valid given the absence of coupling between two adjacent unit cells, \cite{Liu:2019}. 
and the corresponding complex-valued co-polarized reflection coefficients, denoted $\Gamma_{ss/pp}(\theta,\varphi=0)$ and $\Gamma_{ss/pp}(\theta,\varphi=\pi)$ respectively, are known from the corresponding $Z_\text{surf}$ computed in the previous subsections. 

To compute simple expressions for the co- and cross-polarized scattering coefficients for arbitrary $\varphi$, approximations are required due to the inherent complexity of the problem: We assume that the four elements of the reflection coefficient dyadic $[\Gamma]$ can be expressed as a linear combination of $\Gamma_{ss/pp}(\theta,0)$ and $\Gamma_{ss/pp}(\theta,\pi)$, i.e.,
\begin{equation} \label{eq:Gamma_DualPol}
\begin{aligned}
    \Gamma_{ss}(\theta,\varphi) &= a_1(\theta,\varphi) \Gamma_{ss}(\theta,0) + a_2(\theta,\varphi) \Gamma_{ss}(\theta,\pi),\\
    \Gamma_{ps}(\theta,\varphi) &= a_3(\theta,\varphi) \Gamma_{ss}(\theta,0) + a_4(\theta,\varphi) \Gamma_{ss}(\theta,\pi),\\
    \Gamma_{sp}(\theta,\varphi) &= a_5(\theta,\varphi) \Gamma_{pp}(\theta,0) + a_6(\theta,\varphi) \Gamma_{pp}(\theta,\pi),\\
    \Gamma_{pp}(\theta,\varphi) &= a_7(\theta,\varphi) \Gamma_{pp}(\theta,0) + a_8 (\theta,\varphi)\Gamma_{pp}(\theta,\pi),\\
\end{aligned}
\end{equation}
where the scalars $a_1$ to $a_8$ are the weights for each reflection coefficient for the two planes of anisotropy and generally depend both on elevation and azimuth angles. The weights $a_1$ to $a_8$ in this linear system can be determined given measured or simulated data by using various numerical methods such as the least mean square. However, in this work, we further exploit the symmetries in the unit cell (square lattice, orthogonal anisotropy axes) and draw intuition from uniaxial anisotropic materials \cite{Goodman1996,Tret}, in order to produce elegant and simple $\varphi$-only dependent expressions for the weights:
\begin{equation} \label{eq:Gamma_a1a8}
\begin{aligned}
    &a_1 = a_8 \approx \cos^2\varphi,\\
    &a_2 = a_7 \approx \sin^2\varphi,\\ 
    &a_3 = a_5 \approx -\cos\varphi\sin\varphi,\\ 
    &a_4 = a_6 \approx +\cos\varphi\sin\varphi.
\end{aligned}
\end{equation}
These trigonometric expressions are not unique but, by only maintaining the azimuth $\varphi$ dependence, they provide an excellent balance between simplicity and accuracy. We found that the latter remains reasonable as long as the anisotropy between the two planes is weak (i.e., the reflection coefficients in the $\varphi=0$ and $\pi$ planes are not very dissimilar) and the incidence angle is not too oblique, $\theta>\pi/3$.

% ===========================
\section{Results}\label{sec:Results}
% ===========================

%-------------------------------------------------
\subsection{Model Validation in Tunable Perfect Absorption}
% ------------------------------------------------

To demonstrate the usefulness of the developed ECM/TLM we will use it to compute the RC lumped-load values and geometric dimensions required for tunable perfect absorption, and the corresponding reflection spectra, for various $(\theta,\varphi)$ and both polarizations. The ECM/TLM prediction is in all cases compared to FWS spectra, for the same parameters, to validate its accuracy and range.

The cell period is $D^{(1\times1)} = 6.5$~mm and $D^{(2\times 2)} = 6.8$~mm, for the two cases, while the rest of the structural parameters remain the same: $g = 0.7$~mm, $h = 2.2$~mm, $\varepsilon_r = 2.2$, $\tan\delta=0.0009$, $w_L = 0.5$~mm, and $\ell_L=g$. In all of the cases the target operational frequency is set to 5.5~GHz. Naturally, all unit cell designs can be scaled-down for operation at higher frequencies, i.e., in the microwave X-band 10~GHz or millimeter wave band 28~GHz.

\begin{figure}[!tb]
    \centerline{\includegraphics[width=85mm]{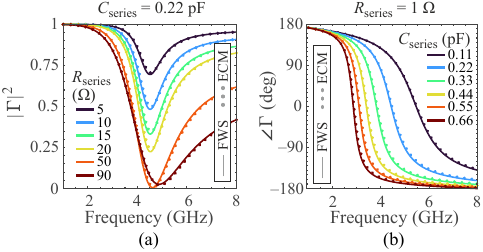}}
    \caption{RC-tunable co-polarized reflection spectra of the $1 \times 1$ cell under normal incidence for varying (a) $R_\text{series}$ or (b) $C_\text{series}$.}
    \label{fig:SpectraExampleNormal}
\end{figure}

We start by showing indicative reflection spectra of the $1\times1$ cell in Fig.~\ref{fig:SpectraExampleNormal} when only the series resistance or only the series capacitance in the lumped load is changed: the former strongly affects the Q-factor and the latter the resonance frequency, respectively. Note the excellent agreement between the ECM/TLM (dotted curves) and the FWS (solid curves).

\begin{figure}[!tb]
    \centerline{\includegraphics[width=85mm]{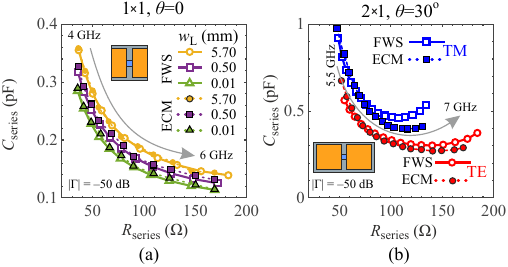}}
    \caption{Optimization of the lumped RC values for $f$-tunable perfect absorption.  (a) $1 \times 1$ cell under normal incidence for three different LFE width, (b) $2 \times 1$ cell under $\theta = 30^\circ$ oblique incidence both polarizations.}
    \label{fig:OptRCvarFreq}
\end{figure}

We then use the ECM/TLM and FWS to compute the series RC loads required for perfect absorption (reflection magnitude lower than $-50$~dB) as the operating frequency is varied, Fig.~\ref{fig:OptRCvarFreq}; we also study the effect of changing the LFE width ($w_L$) or swapping the polarization in mildly oblique incidence. We find good ECM/TLM versus FWS agreement for deviations up to $\pm10\%$ of the central frequency.

\begin{figure*}[!tb]
    \centerline{\includegraphics[width=160mm]{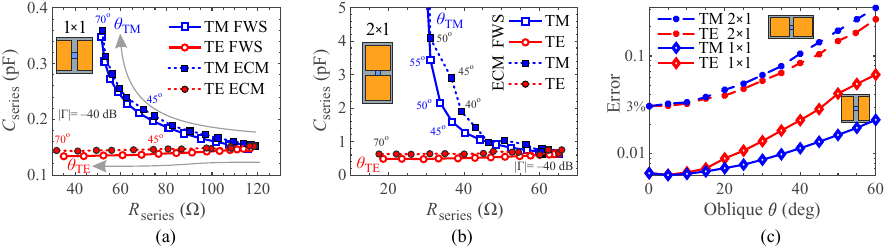}}
    \caption{Optimized lumped RC values for perfect absorption under varying obliquity angle $\theta$ at the frequency of $5.5~\mathrm{GHz}$ for both polarizations in the (a) $1 \times 1$  and (b) $2  \times 1$  cell configuration. (c) Spectrally-weighted error versus obliquity angle $\theta$.}
    \label{fig:OptRCvarTheta}
\end{figure*}

Finally,  we use compute the lumped loads required for perfect absorption for varying oblique angle $\theta$ varies in both polarization planes, at $5.5$~GHz, Fig.~\ref{fig:OptRCvarTheta}. We note the different trends for TE and TM polarizations \cite{Liu:2019} and the good agreement for angles up to $70^\circ$ and $45^\circ$ for the $1\times1$ and $2\times1$ cell, respectively. The spectrally-weighted error in complex $|\Gamma_\text{ECM/TLM}-\Gamma_\text{FWS}|^2$ is depicted in Fig.~\ref{fig:OptRCvarTheta}(c), where the weight is chosen as $W(f)=1-|\Gamma_\text{FWS}(f)|^2$.

%-------------------------------------------------
\subsection{ECM validation for Arbitrary Plane of Incidence}
% ------------------------------------------------

\begin{figure}[!tb]
    \centerline{\includegraphics[width=85mm]{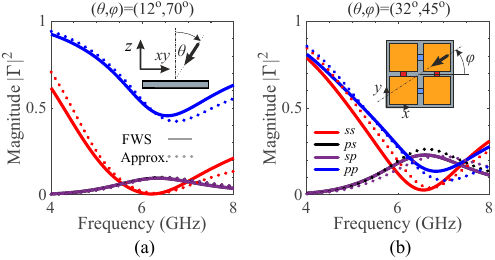}}
    \caption{Spectra of the $2 \times 2$ cell co- and cross-polarized reflection coefficients for two different incidence directions. In both panels, and with reference to Fig.~\ref{fig:Geoms}(d), the lumped RC values are $RC_x=[100~\Omega,1~\text{pF}]$ and $RC_y=[50~\Omega,0.1~\text{pF}]$, respectively.}
    \label{fig:SpectraSkewedPOI}
\end{figure}

The proposed model can be used to compute the co- and cross-polarized reflection spectra under an arbitrary plane of incidence ($\varphi$) and/or oblique incidence ($\theta$), and for different loading in the two axes of the $2\times2$ cell configuration. To demonstrate the accuracy, we compute the four spectra for two arbitrarily chosen incident plane wave directions $(\theta,\varphi)$ and loadings in Fig.~\ref{fig:SpectraSkewedPOI}. 

The comparison between the approximate model (dashed curves) and the FWS (solid curves) is agreeable for all four scattering parameters.%, while the numerically optimized LMS (dash-dot curves) using FWS-fitted $a_{1\rightarrow8}$ weights is also shown. 
The minor discrepancy $\Gamma_{sp}\neq \Gamma_{ps}$ in the model is due to not explicitly enforcing reciprocity in the case of arbitrary incidence plane, which would complicate the expressions for the weights in Eq.\eqref{eq:Gamma_a1a8}. Finally, we stress that the case of TM polarized oblique incidence on a uniaxially anisotropic medium is a known `incomplete' problem, i.e., one cannot fully decompose the E-field in two anisotropy planes.

% ------------------------------------------------
\subsection{Polarization Manipulation}
% ------------------------------------------------

The reconfigurable $2\times2$ cells allow for polarization control metasurfaces \cite{Pitilakis2021,pitilakis2022multifunctional} by tuning capacitance or resistance along the two rectangular axes of anisotropy, i.e., adjusting the phase difference or relative magnitude of orthogonally polarized components, respectively. Using the developed ECM/TLM, we design and analyze two indicative birefringent waveplates for the case of $\varphi = 45^\circ$, i.e., when the two anisotropic axes are equally illuminated, and assuming phase-only manipulation, i.e., $R_x=R_y=0$. Specifically, we used the ECM/TLM to optimize the $C_x$ and $C_y$ capacitances in the $2\times2$ cell, Fig.~\ref{fig:Geoms}(d), for linear-to-circular ($\lambda/2$ waveplate) and linear-to-orthogonal ($\lambda/4$ waveplate) conversion, both at 5.5 GHz; we examined the normal and $\theta = 30^\circ$ oblique incidence cases. The we compared the ECM/TLM versus FWS spectra for these capacitance values.

For ideal linear cross-polarization conversion (half-wave plate), the goal is to achieve $|\Gamma_{ss}|=0$ and $|\Gamma_{ps}|=1$; in practice a magnitude ratio of $\pm20$~dB denotes full conversion. The ECM-computed capacitance values for this condition at 5.5~GHz are $C_x\approx1.23$~pF and $C_y \approx 0.13$~pF for normal incidence and $C_x\approx3.87$~pF and $C_y \approx 0.25$~pF for oblique incidence. For the ideal linear-to-circular (and vice-versa) polarization conversion, we need to implement a quarter-wave plate, i.e., the phase difference between the co- and cross-polarized reflection coefficients must be $\pm 90^\circ$ and their amplitudes must be equal, resulting in an axial ratio (AR) equal to one (0~dB). Using our ECM, we computed that these conditions are met when $C_x \approx 2.51$~pF and $C_y \approx 0.45$~pF for normal incidence and when $C_x \approx 2.95$~pF and $C_y \approx 0.54$~pF for oblique incidence. 

\begin{figure}[]
    \centerline{\includegraphics[width=85mm]{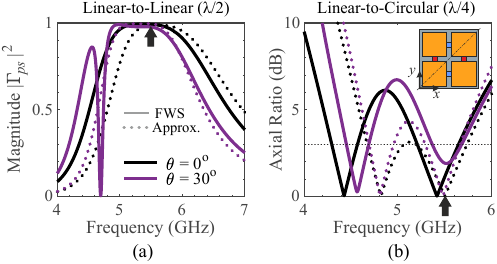}}
    \caption{(a) Magnitude spectra of cross-polarized reflection coefficients for linear cross-polarization conversion. (b) Axial ratio spectra for linear-to-circular polarization. In both panels, the lumped resistance is zero and the incidence plane is at $\varphi = 45^\circ$.}
    \label{fig:PolarizationConversion}
\end{figure}

In Fig. \ref{fig:PolarizationConversion}, we compare the ECM-predicted spectra with the FWS-computed ones, using the same capacitance values. In panel Fig. \ref{fig:PolarizationConversion}(a), for cross-polarized reflection at $5.5$~GHz, we computed $|\Gamma_{ps}| \approx 0.99$ for normal incidence and $|\Gamma_{ps}| \approx 0.98$ for oblique incidence, indicating almost full polarization conversion. The proposed model predicts the 90\% bandwidth with good accuracy in both normal and oblique, but misses the the second resonance that causes the dip at 4.7~GHz (solid purple line). In panel Fig. \ref{fig:PolarizationConversion}(b), for the linear-to-circular polarization conversion, the FWS results at 5.5~GHz show a small deviation from the ECM results for the normal and oblique examples. By examining the axial ratio for the FWS results, we can notice that both cases are close to 0 dB and below 3 dB at the desired frequency depicting agreeable `first-guess' accuracy for the model.

% ====================================================
\section{Conclusions and future prospects}
\label{sec:4:Future}
% ====================================================

This paper introduced an equivalent circuit model (ECM) for square-patch based metasurfaces loaded with tunable lumped elements. The ECM takes into account the physical dimensions of the loads and is extended to arbitrary obliquity and azimuthal plane of incidence. A progressively more complicated family of unit cells, labeled $1 \times 1$, $2 \times 1$, and $2 \times 2$, was studied. The `full' $2 \times 2$ variant allows for reconfigurable multifunctionality, dual/independent polarized operation, and minimized intercell coupling for heterogeneously configured metasurfaces. Comparing the developed ECM against full-wave simulation, we found good accuracy, with resonance frequency errors below $15\%$, provided that patch gaps remain much smaller than the period and that the substrate thickness satisfies $h/(\sqrt{\varepsilon_r}\lambda) \gg 0.1$. This model effectively captures primary resonances, making it suitable for polarization conversion and anomalous reflection, while higher-order resonances can be accounted for with more complicated expressions. In all cases, the usefulness of this model as a rapid preliminary optimization tool was demonstrated, keeping in mind that the lumped-load tunability in RIS can always be used to \textit{a posteriori} compensate minor deviations in frequency or performance.

Practical implementations must consider finite apertures, where physical optics can predict diffraction patterns, provided that nonlocal effects are minimal. Additionally, the physical bulk of the lumped elements placed on the metasurface aperture degrades performance as the frequency increases, suggesting alternative configurations such as placing elements behind the metasurface and using vias for connectivity. The development of more advanced models incorporating these features, i.e., nonlocal effects and intricate connectivity, are important future steps stemming from this work.

% ==============================================
\appendix[Mutual Coupling Impedance] \label{app}
% ==============================================
The coupling between the unloaded patch pair in the $2\times1$ unit cell in Fig.~\ref{fig:Geoms}(c) can be approximated by analyzing the {coupled microstrip waveguide} formed along the $x$-direction. When a plane wave polarized perpendicularly to the patch-gap [$x$-axis in Fig.~\ref{fig:Geoms}(c)] obliquely illuminates the unit cell from above, it excites an {odd mode} propagation along the slotline which then resonates inside each $\ell$-long cell in the $y$-dimension. This transverse resonance contributes to the $Z_\text{cpl}$ term which can be evaluated as
\begin{equation} \label{X4}
    Z_\text{cpl} = \frac{1}{j\omega (C_m \ell)} + j\omega (L_s \ell),
\end{equation}
where $C_m$ and $L_s$ are the unit length capacitance and inductance of the equivalent TL, respectively. The full model comprises a capacitance arising from the proximity of the square patches, an inductance due to the interaction between the patches and the ground plane, a resistance associated with metal losses, and an admittance accounting for dielectric losses. To simplify our ECM, the resistance can be neglected when PEC is used and dielectric losses can be incorporated into the effective permittivity. The mutual capacitance and self inductance can be derived using the formulation in Section 8.5.1 in \cite{gupta}. Given that our unit cells satisfy constraints such as a narrow patch-gap and a thin substrate, the formulas can be simplified by dropping the negligible terms. So, after defining the shorthand variable $C_e$ 
\begin{equation} \label{X7}
    C_e = \frac{1}{2} \biggl( \varepsilon_{r\text{sub}}\varepsilon_0  \frac{w}{h} + \frac{\sqrt{\varepsilon_{r,\text{eff}}}}{c_0 Z_\text{TL,MS}}\biggr),
\end{equation}
we present the simplified formulas for the $C_m$ and $L_s$:
\begin{equation} \label{X5}
    C_m = \frac{2 \varepsilon_0 }{\pi} \Biggl[\varepsilon_{r,\text{inp}} \ln\biggl(\frac{16h}{\pi g}\sinh\Big(\frac{\pi w}{2h}\Big) \biggr) 
    + \ln\biggl(4+8 \frac{w}{g}\biggr)\Biggr] -  C_e,
\end{equation}
\begin{equation} \label{X6}
    L_s = \frac{\mu_0 \varepsilon_0}{C_e};
\end{equation}
in these formulas $\ell\equiv w=w_p$, where $w_p$, $h$ and $g$ are the corresponding dimensions of the $2 \times 1$ cell [see Fig.~\ref{fig:Geoms}(c)], $c_0$ is the speed of light in vacuum, and $Z_\text{TL,MS}$ is the characteristic impedance of the $w_p$-wide microstrip on the same substrate \cite{pozar}.

%=================
%\section*{References}
% ===============

% NOTE: For arXiv LaTeX submission the BBL file is required (not the BIB!)
\bibliographystyle{IEEEtran}
\bibliography{this.bib}

\vfill

\end{document}